# Knowledge transfer in a tourism destination: the effects of a network structure

Rodolfo Baggio[a,*] and Chris Cooper[b]

[a] *Master in Economics and Tourism and Dondena Center for Research on Social Dynamics, Bocconi University, Milan, Italy.*
[b] *The Business School, Oxford Brookes University, UK.*

**Abstract**

Tourism destinations have a necessity to innovate to remain competitive in an increasingly global environment. A pre-requisite for innovation is the understanding of how destinations source, share and use knowledge. This conceptual paper examines the nature of networks and how their analysis can shed light upon the processes of knowledge sharing in destinations as they strive to innovate. The paper conceptualizes destinations as networks of connected organizations, both public and private, each of which can be considered as a destination stakeholder. In network theory they represent the nodes within the system. The paper shows how epidemic diffusion models can act as an analogy for knowledge communication and transfer within a destination network. These models can be combined with other approaches to network analysis to shed light on how destination networks operate, and how they can be optimized with policy intervention to deliver innovative and competitive destinations. The paper closes with a practical tourism example taken from the Italian destination of Elba. Using numerical simulations the case demonstrates how the Elba network can be optimized. Overall this paper demonstrates the considerable utility of network analysis for tourism in delivering destination competitiveness.

**Keywords***:* epidemic diffusion models: knowledge transfer; network analysis; tourism destinations.

*Corresponding author. Email: rodolfo.baggio@unibocconi.it

# Introduction[1]

Knowledge transfer, cultural variables and social embeddedness are key determinants of global competitiveness for advanced regions and nations, and foster a transformation of capitalism towards a 'knowledge economy' (Dayasindhu, 2002; Tödtling et al., 2006; Uzzi, 1996). Tourism is basically a service industry and its management practices are highly focused on the efficiency and effectiveness of the information and knowledge exchanges that happen between the different organizations that need to collaborate to deliver composite products (Otto & Ritchie, 1996). In this respect it can be argued that in the global tourism market it is destinations, not individual businesses, that compete to attract more customers (Ritchie & Crouch, 2003).

---
[1] An earlier version of this paper has been presented at the IASK Advances in Tourism Research 2008 Conference, Aveiro, Portugal, 26-28 May 2008.

In the twenty first century, tourism destinations have an imperative to innovate and remain competitive in an increasingly global competitive environment. A pre-requisite for innovation is the understanding of how destinations source, share and use knowledge. However, the majority of the knowledge management literature and applications are concerned with individual organizations rather than the complex amalgams of organizations that characterize destinations. Of course, the focus on the individual organization can be applied to tourism enterprises, destination management organizations and to government ministries and departments. On the other hand, if knowledge management is to be an effective tool in tourism innovation, then we also need to consider how it can benefit the destination level of organization. This conceptual paper examines the nature of networks and how their analysis can shed light upon the way that destinations can share and benefit from knowledge as they strive to innovate and be competitive. It therefore is intended as a theoretical contribution, introducing the concepts of network analysis, network metrics and epidemic diffusion models and it aims to demonstrate the utility of these approaches for understanding how destinations function.

**Knowledge and Networks**

There are to date, only a small number of examples and applications of knowledge management across destination networks (Baggio, 2007; da Fontoura Costa & Baggio, 2008; Scott et al., 2008a; Scott et al., 2008b). However, recognition of the significance of the approach is growing as practitioners recognize the value of knowledge sharing not just within the organization, but also through networks, and in particular the encouragement of partnerships within destinations. It is characterized by the fact that the early phases of knowledge management were portrayed by the phrase 'knowledge is power'. The new thinking argues, 'sharing is power' and creates 'communities of knowledge' at the destination level.

**Destinations as networks of organizations**

Tourism more than most economic sectors involves the development of formal and informal collaboration, partnerships and networks to deliver the product. In one Australian study, tourism was found to be the economic sector with the most inter-organizational networks (Bickerdyke, 1996). A significant tourism literature on these topics exists in the discussion of partnerships and collaboration (Bramwell & Lane, 2000; Hall, 1999; Selin, 2000; Selin & Chavez, 1995) and networking (Copp & Ivy, 2001; Halme, 2001; Tinsley & Lynch, 2001; Tyler & Dinan, 2001). Indeed one stream of the tourism literature examines tourism (Leiper, 1990), destinations (Carlsen, 1999) and market niches (Scott & Laws, 2004) as a system of interrelated components.

The view of destinations as networks, and more generally as complex dynamic systems (Baggio, 2008b), is amenable to analysis using techniques such as social network analysis. A social network has been defined as a specific set of linkages among a defined set of persons, with the additional property that the characteristics of these linkages as a whole may be used to interpret the social behavior of the persons involved (Mitchell, 1969). Social network analysis delivers a number of useful outcomes. It provides a means of visualizing complex sets of relationships and simplifying them and is therefore useful in promoting effective collaboration within a group, supporting critical junctures in networks that cross functional, hierarchical, or geographic boundaries; and ensuring integration within groups following strategic restructuring initiatives (Cross et al., 2002). In recent times these analysis methods have been applied to the tourism sector and have provided interesting insights (Baggio, 2007; da Fontoura Costa & Baggio et al., 2008a; Scott et al., 2008a; Scott et al., 2008b).

**Destination stakeholders**

A second concept that must be considered in understanding destinations as networks of organizations is that of the stakeholder. The concept is related to changing public sector governance as well as participatory management in the private sector. Stakeholders are the people who matter to a system. A stakeholder is any person, group or institution that has an interest in a development activity, project or program. This definition includes intended beneficiaries and intermediaries, winners and losers, and those involved or excluded from decision-making processes (SDI, 1995).

Stakeholder theory, pioneered by Freeman (1984) suggests that an organization is characterized by its relationships with various groups and individuals, including employees, customers, suppliers, governments, and members of the communities. According to Freeman: "[a] stakeholder in an organization is (by definition) any group or individual who can affect or is affected by the achievement of the organization's objectives" (p. 46).

Thus, a group qualifies as a stakeholder if it has a legitimate interest in aspects of the organization's activities and, thus, according to Freeman, has either the power to affect the firm's performance and/or has a stake in the firm's performance. Hence the concept of a stakeholder is related to the concept of participative government and the growth of community activism. Interestingly, identification and consultation with stakeholders originally started as a means of increasing the effectiveness of business but has come to be seen as a matter of business ethics and principles (Sautter & Leisen, 1999).

In the discussion here, stakeholders are organizations that have some role in the tourism destination. However all stakeholders are not created equal. Stakeholders may be classified both in terms of their individual characteristics as well as their characteristics in relation to networks. A common approach to classifying stakeholders is to do this in terms of key, primary and secondary stakeholders. Stakeholder analysis is a tool which helps an understanding of how operators affect the creation and dissemination of information in a destination and the resultant policies and activities. It is particularly useful in highlighting the challenges that need to be faced to change knowledge management behavior, develop capabilities and tackle problems.

**Networks and knowledge transfer**

Information and knowledge flows in a destination network are relevant mechanisms for the general behavior of the system. Productivity, innovation and economic growth are strongly influenced by these processes, and the way in which the spread occurs can determine the speed by which individual actors perform and plan their future actions at the destination; in other words the structure of the network will be influential in determining the efficiency of the destination's attempts to share knowledge and innovate, and so remain competitive (Argote & Ingram, 2000).

The literature in this field has dealt with two main issues: the mechanisms and the processes of knowledge acquisition within a single stakeholder (e.g. a company, association or group) and the diffusion within the destination network formed by groups of stakeholders, based on their similarity (industrial clusters, for example), or their spatial location. The topology of the destination network formed by the different stakeholders and their formal and informal relationships has proved to be an important determinant when explaining the mechanisms by which ideas, information and knowledge 'travel' from one element of the system to another (Chen & Hicks, 2004; Da Costa & Terhesiu, 2005; López-Pintado, 2004; Valente, 1995).

Social networks are the main channel through which these phenomena unfold. It has been shown in many cases by sociologists and economists that a dense and well formed social network favors a stakeholder's attitude to search for new opportunities and to share experiences, particularly in the presence of dynamic unpredictable environments. This has a beneficial effect on the development of the community in which they are embedded (Inkpen & Tsang, 2005; Levin & Cross, 2004; Vega-Redondo, 2006). As an example, Ingram and Roberts (2000) describe how the intense web of relationships among managers of Sydney hotels has allowed the amalgamation of many best practices, with the result of improving the performance and the profitability of their hotels. Social network analysis tools have thus been used to study such phenomena and have proved to be effective in explaining the general characteristics of networks (Birk, 2005; Cross et al., 2002; Cross et al., 2000).

Many theories have been proposed to describe and explain these diffusion processes. The remainder of this paper is dedicated to a general overview of them and to the exposition and discussion of a simple simulation model.

**Epidemic diffusion models**

The most commonly used models for the flow of knowledge or information through networks are based on an analogy with the diffusion of a disease (Bailey, 1975; Diekmann & Heesterbeek, 2000; Hethcote, 2000). There is clear analogy here between the transmission of disease and the transmission of knowledge through a network. A long tradition of epidemiology studies has dealt with the issue of describing the spread of a disease in a population of living organisms. From Daniel Bernoulli's (1760) analysis of smallpox at the end of 18th century, mathematical modeling and numerical simulations have helped in the study of the effects of bacterial, parasitic and viral pathogens, infections and the possible countermeasures.

The mathematical models used are based on the cycle of infection in an individual. The 'host' is first considered susceptible (S) to the disease. Then, if exposed to the infection it becomes infected (I) and is considered infectious for a certain period of time. Finally, the individual can recover (R) by acquiring some immunity or by being 'removed' from the population. These basic elements (along with some possible variations) are used to characterize the different models which are identified by the initials of the types of infection considered. Therefore, we have SI models, in which hosts can be only susceptible or infected; SIS models in which they go through a complete cycle: susceptible, infected, then susceptible again; and SIR models which consider susceptible individuals that are infected and end their process by being removed (i.e. immunized or eliminated from the initial population). Again the analogy with knowledge flow though a destination network is clear - stakeholders may be susceptible to receiving new knowledge, but until they are 'infected' knowledge transfer does not take place.

The mathematical treatment has much in common with the one used to describe the percolation phenomenon (the diffusion of a fluid through a porous medium). The curves describing the results of the infection are mostly s-shaped curves belonging to the family of logistic curves, and are in many cases similar to the ones representing the growth of a population. Traditionally all epidemic models have assumed perfect mixing: i.e., all individuals are equally able to infect all others, and have taken into account a random distribution of the contacts between individuals that are responsible for the infection (diseases spread through some kind of contact between the population elements). In some cases the models are refined by making assumptions about the population affected: e.g., the way the hosts react to the infection, recover from the disease or are removed from the population.

Hosts in a population can be represented by the nodes of a network in which the contacts constitute the links. Recent advances in the study of complex networks have allowed a reconsideration of epidemic diffusion models to take into account the effects of non-homogeneous network topologies (Kuperman & Abramson, 2001; Pastor-Satorras & Vespignani, 2001, 2003). These effects are quite important. For example, it has been known for a long time (Kermack & McKendrick, 1927) that the SIS model shows a clearly defined threshold condition for the spread of an infection. This threshold depends on the density of the connections between the different elements of the network. However, this condition is valid only if the link distribution is of a random nature, while in some of the structured, non-homogeneous networks that make up the majority of real systems, this threshold does not exist. Once initiated, the diffusion process unfolds over the whole network (Pastor-Satorras & Vespignani, 2001).

The formulation of an epidemiological model leads to the layout of a system of differential equations which can be demanding. In the last few years however, the availability of computational tools (both hardware and software) has fostered the development and the usage of numeric simulation models. In what follows we shall use this approach to analyze a tourism case taken from Italy.

**Network models**

A long tradition, prompted by the 1736 paper by Leonhard Euler on the Königsberg bridges, has provided a widespread set of mathematical tools for analyzing networks and the graphs they represent. During the 20th century, the ideas and techniques developed for the study of these abstract objects have been applied to several fields. In particular, realizing that a group of individuals can be represented by enumerating the stakeholders of the group and their mutual relationships, sociologists have used some of the methods belonging to graph theory to study their patterns of social relations (Freeman, 2004; Wasserman & Faust, 1994). Furthermore, in the last decade, the community of physicists and mathematicians has exploited the vast amount of data available through the Internet to develop a whole new set of models. With these it has been possible to describe the static, structural and dynamic characteristics of a wide range of both natural and artificial complex networks (Boccaletti et al., 2006; da Fontoura Costa et al., 2008; Watts, 2004).

The shape of a network and the relationship between its nodes, the network topology, has been found to be a crucial determinant of the functions the system performs, and of the quality of the communication between nodes. The literature on complex networks proposes a number of different measurements with which it is possible to characterize the network topology. Their calculation derives mainly from the work done by mathematicians in graph theory and is based on a matrix representation of the links between the network nodes (Bollobás, 1998; Godsyl & Royle, 2001). The rich set of metrics used today originates from the combination of those coming from the tradition of social network analysis with the outcomes of the recent work (Boccaletti et al., 2006; Börner et al., 2007; Bornholdt & Schuster, 2002). Some of these are have been recognized to be the most important to fully characterize topology and behaviors of a complex network and can be applied to destinations (Baggio et al., 2008a):

- *degree distribution* $P(k)$: the statistical distribution of the number (and sometimes the type) of the linkages among the network elements;
- *average path length* $L$: the mean distance between any two nodes and diameter $D$: the maximal shortest path connecting any two nodes. Small values for $D$ and $L$ indicate good compactness of a network; at least, of its main connected component (i.e. disregarding isolated nodes);

- *clustering coefficient* C: representing the concentration of the connection of the node's neighbors in a graph and giving a measure of the heterogeneity of local density of links; and
- *efficiency* (at a local $E_{loc}$ or global $E_{glob}$ level): which can be interpreted as a measure of the capability of the system to exchange information over the network; and assortativity: which gauges the correlation between the degrees of neighbor nodes. If positive, the network is said to be assortative. In such a network, well-connected elements tend to be linked to each other. This quantity, related to the clustering coefficient, has been recently shown to influence directly the formation of strongly connected sub-networks or communities and to give an indication of their strength (Quayle et al., 2006).

The mathematical expressions for these quantities can be found in one of the recently published reviews of the research in this area (Caldarelli, 2007; da Fontoura Costa et al., 2007).

**Computer simulations**

In addition to describing and explaining phenomena, numerical simulations allow experiments to be performed in fields where these would not otherwise be feasible for both theoretical and practical reasons. A network is a system which may comprise a very large number of elements and its topological characteristics have a direct relationship with many dynamic processes. It would be therefore be interesting to experiment with different configurations to measure these effects in order to better understand how these differing configurations influence the behavior of the whole destination system.
Social scientists have long used simulation techniques (Inbar & Stoll, 1972). The wide availability of computing power and of efficient programming languages, coupled with a much simpler access to data has, in recent decades, greatly enlarged the amount of attention given to these methods and their practical uses (Castellano et al., 2009; Conte et al., 1997; Gilbert, 1999; Suleiman et al., 2000). A widely used environment to perform simulations is the series of toolkits developed to implement agent-based models (ABM). The idea of such simulations is that a system is composed of a number of entities (agents) which behave according to some simple rule (Flake, 1998; Wolfram, 2002). The interactions of the agents can generate some global system property which can then be studied. Variations in the basic rules or in the typology of the agents produce different final configurations for the system. The reliability and credibility of these techniques is generally considered good, provided some basic requirements are met: as recognized in the literature, the most important being the usage of a solid conceptual model and the connection with the particular circumstances for which the simulations are run. In other words: no absolute value can be given to such processes, as their value will be dependent on the specific situation or the specific purpose (Küppers & Lenhard, 2005; Law & Kelton, 2000; Schmid, 2005). With these caveats, these models have proved to be both effective and efficient in reproducing different types of social and natural systems and may be considered a valuable aid in decision making (Tesfatsion & Judd, 2006; Toroczkai & Eubank, 2005). A number of dedicated programs have been developed to help with ABM simulations, and specialized software packages provide libraries with functionalities at different levels of complexity.

**Materials and methods**

Based on the above discussion, we can consider the diffusion of knowledge in a tourism destination as an 'infection' process in which knowledgeable individuals (in our case the destination stakeholders) transfer their knowledge to the other members of the social group with which they have contact. Configuration of the network and the nature of the stakeholders would be expected to influence the efficiency of this process and thus, ultimately, the destination's ability to innovate and be competitive.

**The destination network of Elba, Italy**

The island of Elba, Italy, is part of the Tuscany Archipelago National Park and the third Italian island. It is an important environmental resource owing to its geographic position, temperate climate and the variety and beauty of its landscapes, coast and sea. It is a sea, sport and culture destination, with almost 500,000 tourist arrivals, 3 million overnights per year and several hundred accommodation establishments. Elba is considered a 'mature' tourism destination (Pechlaner et al., 2003; Tallinucci & Testa, 2006) with a long history and which has gone through a number of different expansion and reorganization cycles. The great majority of the stakeholders are small and medium sized companies (SMEs), mostly family-run. Several associations and consortia operate on the island in an attempt to overcome the excessive 'independence' of SMEs by suggesting and developing different kinds of collaboration programs.

The destination network was assembled in the following way. The core tourism companies, organizations and associations operating at Elba are the vertices of a network whose ties are the relationships among them. According to the local tourism board, the list of companies comprises 1028 elements and the connections represent 'business' relations between organizations. They were collected by consulting publicly available sources such as listings of the members of associations, members of management boards, catalogs of travel agencies, marketing leaflets and brochures, and official corporate records (to assess memberships of industrial groups). These data were then verified with a series of in-depth interviews with 'knowledgeable informants' including the director of the tourism board, directors of associations, and tourism consultants. This triangulation (Olsen, 2004) allowed the assessment of the validity of the collected linkages and revealed others. With these additions the network can be reasonably estimated to be almost 90% complete.

Table 1 The main metrics calculated for the whole Elba network and its main connected component

| Metric | Whole network | Connected component |
|---|---|---|
| Number of nodes | 1028 | 627 |
| Number of edges | 1642 | 1642 |
| Density | 0.003 | 0.008 |
| Disconnected nodes | 37% | --- |
| Diameter | 8 | 8 |
| Average path length | 3.16 | 3.16 |
| Clustering coefficient | 0.050 | 0.08 |
| Average degree | 3.19 | 5.21 |
| Average closeness | 0.121 | 0.326 |
| Average betweenness | 0.001 | 0.003 |
| Global efficiency | 0.131 | 0.353 |
| Local efficiency | 0.062 | 0.102 |
| Assortativity coefficient | -0.164±0.022 | -0.175±0.024 |

Complex network analysis techniques were used to calculate the topological characteristics of the system (for definitions and formulas see for example da Fontoura Costa et al., 2007). The metrics were calculated by using available software packages (Pajek, Ucinet) complemented by some Matlab programs developed by one of the authors. Table 1 shows the values calculated for the whole network and for its main connected component (i.e. disregarding isolated nodes). The degree distribution has a power-law behavior $P(k) \sim k^{-\gamma}$ where $\gamma = 2.32 \pm 0.269$ (the scaling exponent is calculated according to Clauset et al., 2007).

The analysis of the main topological characteristics of the Elba network can be summarized as follows:
- the network shows a scale-free topology (power-law behavior of the degree distribution) which is consistent with that generally ascribed to many artificial and natural complex networks;
- the general connectivity is very low (link density) with a very large proportion of disconnected elements; and
- clustering is quite limited, as is the efficiency, both at a local and global level.

These results provide quantitative evidence in favor of recognizing that the 'community' of Elban tourism operators is fragmented in nature. There appears little incentive to group or cluster in a cooperative or collaborative manner as evidenced by considering the clustering and assortativity characteristics. These conditions are also problematic for an efficient flow of information and knowledge through the social system, and this may affect its capabilities to innovate and be competitive in the future. These considerations are in general agreement with previous studies performed by using more traditional qualitative techniques (Pechlaner et al., 2003; Tallinucci & Testa, 2006).

**Simulating knowledge flow in the Elba network**

The Elba network can be used to perform a simulation of the transfer of information and knowledge across the network. The objective here is to assess the present situation and to test the capability of the destination in absorbing the knowledge transferred when changing some of its structural parameters.

In our simulation a simple SI epidemiological model is used. Despite its simplicity, this class of model has been shown to be quite effective and to be a good approximation of more refined and complex models (Barthélemy et al., 2005; Xu et al., 2007). In addition, it is suitable for describing the knowledge transfer process. In fact, we may well reasonably assume that once knowledge has been transferred to a new host, it will retain the knowledge received, therefore it will remain 'infected'. This is an essential pre-requisite to innovation as unless the knowledge is transferred and used by enterprises at the destination, innovation will not occur.

The algorithm used for the simulation is the following:
1. the network is loaded;
2. one randomly chosen stakeholder starts the spread by infecting a proportion $p_i$ of its immediate neighbors. In tourism, this stakeholder is often a government-funded tourist board or economic development agency;
3. at each time step the infected elements transfer the knowledge to a proportion $p_i$ of their immediate neighbors; and
4. the process ends when all the network nodes have been infected.

As a parameter for the model, the capacity of the solitary stakeholders to transfer knowledge is used. It can be expressed as a probability $p_i$, whose value controls the number of neighbors which are informed by a single stakeholder. This accounts for an important difference between information and knowledge flows and the spread of viruses. While viruses tend to be indiscriminate, infecting any susceptible individual, knowledge is selective and is passed by its host only to a limited set of the individuals with which it has relations (Huberman & Adamic, 2004). Moreover, particular actors can have difficulties in acquiring and retaining all the knowledge available to them (a feature usually called absorptive capacity, see for example Cohen & Levinthal, 1990; Priestley & Samaddar, 2007) due to their internal functioning or because of the associated costs. In tourism, this issue of absorptive capacity is critical, particularly given the dominance of SMEs in the sector.

We can assume that the capacity of transferring knowledge is different for the different 'sizes' of companies involved. Therefore, the network nodes have been divided into three categories: large, medium and small. In our case we have the following proportions: large = 7%, medium = 16%, small = 77%. The values for the proportion of neighbors informed used in the simulation runs are (arbitrarily) set as: $p_{large} = 1$, $p_{medium} = 0.8$, and $p_{small} = 0.6$. Since the structural characteristics of the network, and particularly the cohesion among stakeholders, can be a factor influencing the knowledge transfer process, the experiment has also been performed with a modified version of the original network (Levin & Cross, 2004; Reagans & McEvily, 2003). This has been obtained by rewiring the connections while leaving unchanged the original connectivity (i.e. the number of immediate neighbors of each stakeholder and overall density of linkages), in order to obtain higher local efficiency and clustering coefficient. The algorithm used is similar to the one proposed by Maslov (2002). For all the simulations only the main connected component of the network has been considered.

The new network has a clustering coefficient $C = 0.274$ and mean local efficiency $e_{loc} = 0.334$, as opposed to the original one whose values are $C = 0.084$ and $e_{loc} = 0.104$ (see Table 1, connected component). It should be noticed that both values are lower than those reported by the literature for social networks (Dorogovtsev & Mendes, 2002; Watts, 2004).

A synthetic network of the same size and order (same number of nodes and links) but with a random distribution of links is used as a comparison in this case. The model has been implemented with Netlogo (Wilensky, 1999) and is a derivation of some of the distribution library models (Rumor Mill as modified by F. Stonedahl http://www.cs.northwestern.edu/~fjs750/netlogo/).

Given the conceptual and theoretical nature of the work presented in this paper, the choice of all the parameters used in the model, although 'reasonable', is by some means arbitrary. However, they are non influential for the aims of the simulations, their choice was determined by trading off statistical significance of the outcomes and minimization of computational efforts.

**Results and discussion**

The simulation results are shown in Figure 1 and Figure 2. Figure 1 depicts the cumulative number (as a percentage of total) of stakeholders that are 'infected' as function of time. Fig. 2 is the differential version, i.e. the number of informed actors at each time interval.

As can be seen, a random homogeneous network (Rnd) shows the slowest diffusion process with respect to any other network whose connections have a structured non-homogeneous distribution. This comparison with a 'random' network reinforces the idea that the structure of the social network has a noticeable impact on the phenomenon studied. The first series of simulations (EDiff vs. EN) highlight a difference in speed. It looks as if removing the differences in the capability of tourism stakeholders to transfer knowledge to other members of the community can improve, in a visible way, the whole diffusion process.

The 'topology' effect described above is much more evident in the second series of simulations (RW). In this case the model has been used by changing the structure of the actual network. The runs are based on the rewired network having a much greater clustering coefficient, i.e. a much greater degree of local cohesion among the tourism stakeholders.

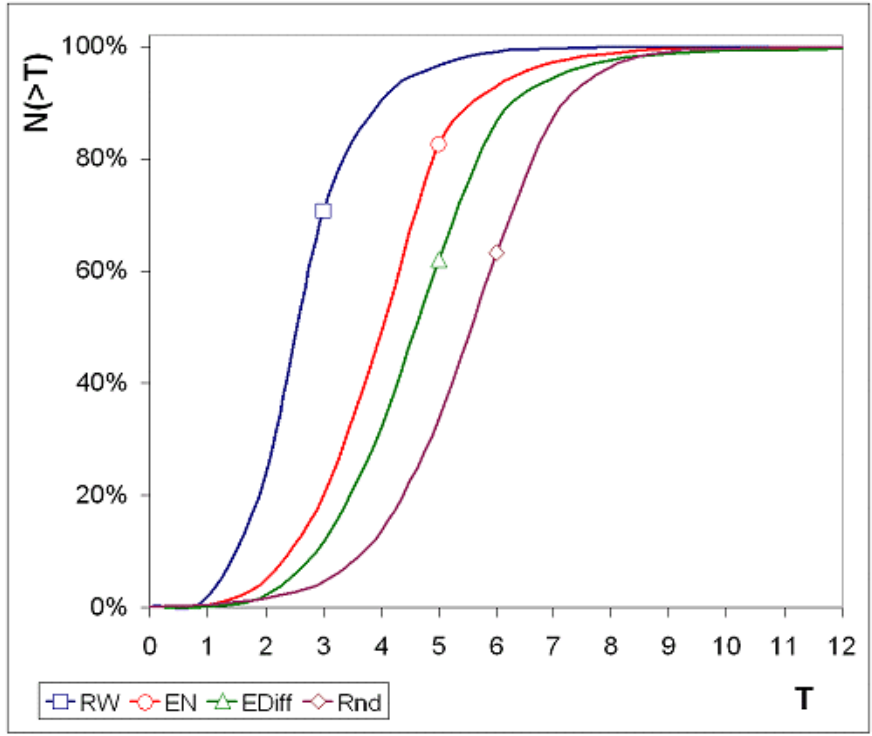

Figure 1. Cumulative percentage of informed stakeholders for the simulations performed: rewired network (RW), Elba network with equal probability of transmission (EN), with probabilities scaled according to stakeholder size (EDiff) and a network of same size with a random distribution of links (Rnd). Curves are averaged over 10 realizations of the simulations.

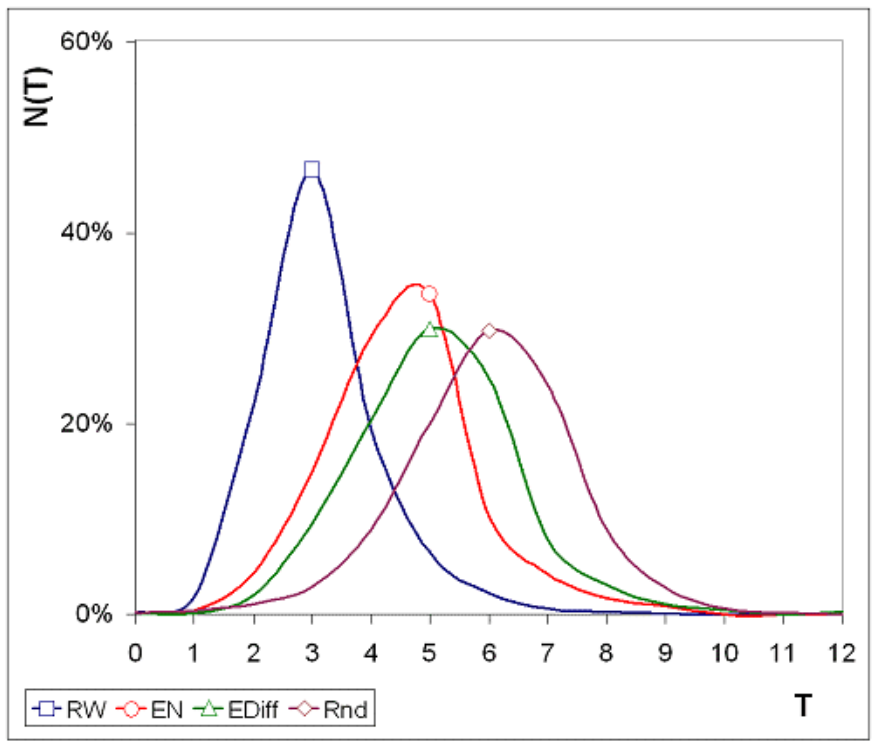

Figure 2. Differential curve of knowledge spreading for the simulations performed: rewired network (RW), Elba network with equal probability of transmission (EN),

with probabilities scaled according to stakeholder size (EDiff) and a network of same size with a random distribution of links (Rnd). Curves are averaged over 10 realizations of the simulations.

In conclusion, the simulations give us a clearly identifiable pattern. The knowledge diffusion process is faster in the case of a structured network (e.g. the power-law distributed Elba network) than in a random one. A much higher improvement is found when considering the increase in clustering. Table 2 summarizes these differences by showing the peak time of the diffusion process and the percentage differences.

Table 2 Time of peak diffusion ($T_{peak}$) and speed improvements in diffusion over different network topologies (Rnd = random network, EDiff = differential actors capabilities, EN = original Elba network, RW = rewired network).

| Network | $T_{peak}$ | Speed improvement |
|---|---|---|
| Rnd | 6.1 | ---- |
| EDiff | 5.2 | 16% |
| EN | 4.8 | 22% |
| RW | 2.9 | 52% |

We must therefore deduce that a very important determinant for the spread of knowledge in a socio-economic system such as a tourism destination is the presence of a structured topology in the network of relations that connect the different stakeholders. Moreover, the existence of a well-identified degree of local cohesion strongly influences this process. This supports the notion that destination stakeholders should be encouraged to form clusters and to both compete and cooperate to raise the overall competitiveness of the destination. Often the public sector intervenes to initiate such cooperative processes, given the combative nature of SMEs. However, public sector support can facilitate a network and provide ongoing support, but it is the destination stakeholders who must operate the network.

These results are not completely new. The effect is the one identified by Granovetter (1973; 1983) as the strength of weak ties and reconfirmed by the more recent works on the so-called small world networks (Latora & Marchiori, 2001; Uzzi & Spiro, 2005; Watts & Strogatz, 1998). Moreover, several authors have empirically found this behavior (Reagans & McEvily, 2003; Sorenson et al., 2006). Here, for the first time, a tourism destination is used as test case.

**Concluding remarks**

This paper has outlined a theoretical approach to analyzing the functioning of tourist destinations. In doing so it has demonstrated the benefits of importing analytical and theoretical techniques of network analysis to tourism destinations. By coupling these theoretical and analytical approaches with a thorough understanding of the destination and its stakeholders, we can diagnose the efficiency of the destination's network structure and its implications for competitiveness. We can also begin to utilize policy instruments to intervene and to make the network more efficient. In other words, in a case such as Elba, the simulations can be used to create development scenarios in which the efforts to move towards strong forms of collaboration are increased, even if at a very 'local' level. This can be highly beneficial not only for the stakeholders involved, but for the whole destination.

The approach conceptualizes destinations as networks of organizations and communities where the destination functions by the movement of resources, such as information or investment, through the network. The benefits of conceptualizing destinations in this way are clear. In a knowledge economy, destinations have to innovate to remain competitive. The management of knowledge underpins this innovation and so, an understanding of how knowledge can be managed across complex network organizations is fundamental to this process. For tourism, as has been seen, a particular concern is the fact that most destinations are comprised of SMEs, organizations which tend to be knowledge averse and therefore public sector intervention is needed to establish cooperative frameworks and networks at the destination level. In other words, the theoretical interest in understanding the processes of knowledge transfer in a complex system such as a tourism destination is crucial from the point of view of practitioners.

This implies a future research agenda focused upon network configuration and metrics linked to the competitive performance of destinations. Comparative analysis of destination networks would deliver an understanding of the most effective configurations of destinations. Here, diagnostics can be used to improve connectivity by intervening to mend broken links, or reconfiguring the network to be more efficient. This paper has shown that this is possible by making an initial attempt in applying metrics to destinations. The methods and the techniques used have shown that, once accepted, important 'network' framework results can be derived by studying a specific system. The basic analytical tools allow an assessment of the peculiar characteristics of the structure and functioning of a destination. Computerized numerical simulation models based on the theories of a network can deliver differing development scenarios and show how the system would evolve. It should be observed, however, that the quantitative tools and methods used here are not fully sufficient to provide a full range of results. To move the research agenda forward, knowledge of the specific destination under study combined with qualitative assessments of the sector and local policy can greatly add to the toolbox available to tourism scholars and practitioners, and in turn, better equip them in their effort to understand the complex systems that are networked tourism destinations.